\documentclass[prb,twocolumn,showpacs,floatfix,superscriptaddress]{revtex4-2}

\usepackage{amssymb,amsmath,amstext}    
\usepackage{graphicx} 
\usepackage{epstopdf}
\usepackage{color}  
\usepackage{bm}
\usepackage{appendix}
\usepackage[utf8]{inputenc}
\usepackage{bbold}
\usepackage{bbm}
\usepackage{latexsym}
\usepackage{mathrsfs}
\usepackage[colorlinks=true,citecolor=blue,linkcolor=magenta]{hyperref}

\def\be{\begin{equation}}
\def\ee{\end{equation}}

\def\ba{\begin{eqnarray}}
\def\ea{\end{eqnarray}}
\newcommand*{\defeq}{\stackrel{\text{def}}{=}}

\begin{document}
\title{Tensor network study of the $m=1/2$ magnetization plateau \\ in the Shastry-Sutherland model at finite temperature}
\author{Piotr Czarnik}
\affiliation{Theoretical Division, Los Alamos National Laboratory, Los Alamos, NM 87545, USA}

\author{Marek M. Rams}
\affiliation{Jagiellonian University, Institute of Theoretical Physics, {\L}ojasiewicza 11, PL-30348 Krak\'ow, Poland}

\author{ Philippe Corboz }
\affiliation{Institute for Theoretical Physics and Delta Institute for Theoretical Physics, 
             University of Amsterdam, Science Park 904, 1098 XH Amsterdam, The Netherlands}           

\author{Jacek Dziarmaga}
\affiliation{Jagiellonian University, Institute of Theoretical Physics, {\L}ojasiewicza 11, PL-30348 Krak\'ow, Poland}

\begin{abstract}
  The two-dimensional infinite projected entangled pair states tensor network is evolved in imaginary time with the full update~(FU) algorithm to simulate the Shastry-Sutherland model in a magnetic field at finite temperature directly in the thermodynamic limit. We focus on the phase transition into the $m=1/2$ magnetization plateau, which was observed in experiments on SrCu$_2$(BO$_3$)$_2$. For the largest simulated bond dimension, the early evolution in the high-temperature regime is simulated with the simple update~(SU) scheme and then, as the correlation length increases, continued with the FU scheme towards the critical regime. We apply a small symmetry-breaking bias field and then extrapolate towards  zero bias using a simple scaling theory in the bias field. The combined SU + FU scheme provides an accurate estimate of the critical temperature, even though the results could not be fully converged in the bond dimension in the vicinity of the transition. The critical temperature estimate is improved with a generalized scaling theory that combines two divergent length scales: One due to the bias, and the other due to the finite bond dimension. The obtained results are consistent with the transition being in the universality class of the two-dimensional classical Ising model. The estimated critical temperature is $3.5(2)$~K, which is well above the temperature $2.1$~K used in the experiments. 
\end{abstract}

\maketitle

\section{Introduction}
\label{sec:introduction}

Weakly entangled quantum states constitute a small corner in an exponentially large Hilbert space but are ubiquitous as stationary (ground or thermal) states appearing in condensed-matter physics.
They can be efficiently represented by tensor networks~\cite{Verstraete_review_08,Orus_review_14}, including the one-dimensional (1D) matrix product state (MPS)~\cite{fannes1992}, its two-dimensional (2D) generalization known as a projected entangled pair state (PEPS)~\cite{verstraete2004}, or a multi-scale entanglement renormalization ansatz~\cite{Vidal_MERA_07,Vidal_MERA_08,Evenbly_branchMERA_14,Evenbly_branchMERAarea_14}. The MPS ansatz provides a compact representation of ground states of 1D gapped local Hamiltonians~\cite{Verstraete_review_08,Hastings_GSarealaw_07,Schuch_MPSapprox_08} and purifications of their thermal states~\cite{Barthel_1DTMPSapprox_17}. It is also the ansatz underlying the density matrix renormalization group (DMRG)~\cite{White_DMRG_92, White_DMRG_93,Schollwock_review_05,Schollwock_review_11}. Analogously, the 2D PEPS is expected to represent ground states of 2D gapped local Hamiltonians~\cite{Verstraete_review_08,Orus_review_14} and their thermal states~\cite{Wolf_Tarealaw_08,Molnar_TPEPSapprox_15}, although representability of area-law states, in general, was shown to have its limitations~\cite{Eisert_TNapprox_16}. Tensor networks do not suffer from the notorious sign problem plaguing quantum Monte Carlo methods. Consequently, they can deal with fermionic systems~\cite{Corboz_fMERA_10,Eisert_fMERA_09,Corboz_fMERA_09,Barthel_fTN_09,Gu_fTN_10} as was shown for both finite~\cite{Cirac_fPEPS_10} and infinite PEPS~\cite{Corboz_fiPEPS_10,Corboz_stripes_11}.

The PEPS was originally proposed as an ansatz for ground states of finite systems~\cite{Verstraete_PEPS_04, Murg_finitePEPS_07}, generalizing earlier attempts to construct trial wave functions for specific models~\cite{Nishino_2DvarTN_04}. The subsequent development of efficient numerical methods for infinite PEPS (iPEPS)~\cite{Cirac_iPEPS_08,Xiang_SU_08,Gu_TERG_08,Orus_CTM_09} promoted it as one of the methods of choice for strongly correlated systems in 2D. Its power was demonstrated, e.g., by a solution of the long-standing magnetization plateaus problem in the highly frustrated compound $\textrm{SrCu}_2(\textrm{BO}_3)_2$~\cite{matsuda13,corboz14_shastry}, establishing the striped nature of the ground state of the doped 2D Hubbard model~\cite{Simons_Hubb_17} and new evidence supporting the gapless spin liquid in the kagome Heisenberg antiferromagnet~\cite{Xinag_kagome_17}. Recent developments in iPEPS optimization~\cite{fu,Corboz_varopt_16,Vanderstraeten_varopt_16}, contraction~\cite{Fishman_FPCTM_17,Xie_PEPScontr_17}, energy extrapolations~\cite{Corboz_Eextrap_16}, and universality-class estimation~\cite{Corboz_FCLS_18,Rader_FCLS_18,Rams_xiD_18} pave the way towards even more complicated problems, including simulation of thermal states~\cite{Czarnik_evproj_12,Czarnik_fevproj_14,Czarnik_SCevproj_15, Czarnik_compass_16,Czarnik_VTNR_15,Czarnik_fVTNR_16,Czarnik_eg_17,Dai_fidelity_17,CzarnikDziarmagaCorboz,czarnik19b,Orus_SUfiniteT_18,CzarnikKH,wietek19,jimenez20,poilblanc20}, mixed states of open systems~\cite{Kshetrimayum_diss_17,CzarnikDziarmagaCorboz}, excited states~\cite{Vanderstraeten_tangentPEPS_15,ExcitationCorboz}, or real-time evolution~\cite{CzarnikDziarmagaCorboz,HubigCirac,tJholeHubig,Abendschein08,SUlocalization,SUtimecrystal}.

In parallel with iPEPS, there is continuous progress in simulating systems on cylinders of finite width using DMRG. This numerically highly stable method that is now routinely used to investigate 2D ground states~\cite{Simons_Hubb_17,CincioVidal} was applied also to thermal states on a cylinder~\cite{Stoudenmire_2DMETTS_17,Weichselbaum_Tdec_18,WeichselbaumTriangular,WeichselbaumBenchmark,chen20}. However, the exponential growth of the bond dimension limits the cylinder's width to a few lattice sites. Among alternative approaches are direct contraction and renormalization of a three-dimensional tensor network representing a 2D thermal density matrix \cite{Li_LTRG_11,Xie_HOSRG_12,Ran_ODTNS_12,Ran_NCD_13,Ran_THAFstar_18,Su_THAFoctakagome_17,Su_THAFkagome_17,Ran_Tembedding_18}.

In this article, we apply the recent iPEPS finite-temperature (imaginary-time evolution) algorithm from Ref.~\onlinecite{CzarnikDziarmagaCorboz} to a challenging frustrated spin system: the Shastry-Sutherland model (SSM)~\cite{Shastry81} in a magnetic field. It is an effective model of SrCu$_2$(BO$_3$)$_2$~\cite{Kageyama99,Miyahara99,Miyahara03} for which experiments have revealed an intriguing sequence of magnetization plateaus~\cite{Kageyama99,Onizuka00,kageyama00,Kodama02,takigawa04,levy08,Sebastian08,Jaime12,takigawa13,matsuda13,haravifard16,shi19}. On the theory side, much progress has been made in understanding the spin structures realized in these  plateaus~\cite{Miyahara99,momoi00a,Momoi00,fukumoto00b,fukumoto01,Miyahara03,miyahara03b,Dorier08,Abendschein08,takigawa10,manmana11b,Nemec12,Lou12, takigawa13,matsuda13,corboz14_shastry,schneider16,shi19}. Wheres at large magnetic fields the spin structures can be understood as crystals of triplets, at low fields they correspond to crystals of triplet bound states~\cite{corboz14_shastry}. In experiments using ultrahigh magnetic fields up to 118~T~\cite{matsuda13}, a $m=1/2$ plateau was found at low temperatures ($2.1$~K), which has also been predicted in theoretical studies at zero temperature~\cite{Miyahara99, Momoi00, Abendschein08,Jaime12,Lou12,matsuda13}. However, accurate studies at finite temperatures and finite magnetic fields have so far been lacking (for recent works at zero magnetic field, see Refs.~\cite{wietek19,jimenez20}).  In particular, the negative sign problem puts it out of reach of quantum Monte Carlo. In this paper, we use iPEPS to study the finite temperature phase transition into the $m=1/2$ plateau phase to accurately determine the critical temperature and to confirm that the transition belongs to the 2D Ising universality class. 

This paper is organized as follows. We first introduce the SSM in Sec.~\ref{sec:model}. We follow in Sec.~\ref{sec:iPEPS} with a summary of the finite-temperature simple update (SU) and full update (FU) algorithms of Ref.~\onlinecite{CzarnikDziarmagaCorboz} used in this paper. In the same section, we introduce a hybrid SU + FU algorithm that provides better stability in the case of a large bond dimension. In Sec.~\ref{sec:simple}, we summarize a simple scaling theory~\cite{CzarnikDziarmagaCorboz,CzarnikKH} that allows extrapolation to a zero symmetry-breaking field for results that are converged in the bond dimension. This simple theory is applied to the results of our simulations in Sec.~\ref{sec:results}, providing estimates of the critical temperature and evidence for the 2D Ising universality class of the transition. As the obtained numerical results are not fully converged in the bond dimension, in Sec.~\ref{sec:general} we introduce a generalized theory that enables extrapolation to the zero-field limit for results that are close to convergence. The theory yields the critical temperature consistent with the estimates of the simple theory applied to the largest simulated bond dimension. Finally, we conclude in Sec.~\ref{sec:summary}. Further technical details can be found in a series of appendices. In Appendix~\ref{app:SU+FU} we discuss technical details of the simulations with the hybrid SU + FU algorithm, and in Appendix~\ref{app:SU} we compare the FU and SU + FU results to ones obtained with the SU evolution carried on all the way to the critical regime. Effects of the finite environmental bond dimension $\chi$ and Trotter step $d\beta$ are analyzed in Appendix~\ref{app:chidbeta}. Finally, we comment on the U(1) symmetry sectors appearing in our simulations in Appendix~\ref{app:sectors}.

\section{Shastry-Sutherland model}
\label{sec:model}

\begin{figure}[t]
\includegraphics[width=0.8\columnwidth,clip=true]{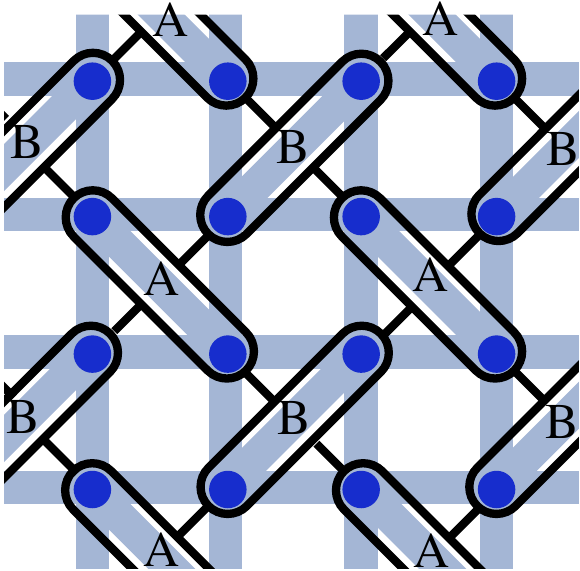}
\caption{
The Shastry-Sunderland model of spins-$1/2$ (represented by blue dots) arranged on a square lattice.
The light blue bands indicate antiferromagnetic Heisenberg couplings between nearest- and some next-nearest-neighbor sites with coupling strengths $J'$ and $J$, respectively.
It can be represented as a checkerboard like lattice of dimers (indicated with black ovals). We label dimers belonging to the two sublattices as $A$ and $B$.
In our simulations, each dimer is combined into a single effective lattice site with a physical dimension $d=4$. This way, we obtain a nearest-neighbor Hamiltonian on an effective square lattice of dimers. The corresponding tensor network ansatz, with one tensor per dimer, is shown with the black lines that indicate the virtual iPEPS bonds (the physical legs of the tensors are omitted here).
}
\label{fig:SS}
\end{figure}

The SSM~\cite{Shastry81} with an external magnetic field is given by the Hamiltonian
\be
H_0 = J' \sum_{\langle i,j \rangle}\mathbf{S}_i \cdot \mathbf{S}_j + J \sum_{\langle \langle i,j\rangle\rangle}\mathbf{S}_i \cdot \mathbf{S}_j - h \sum_i S_i^z,
\label{H}
\ee
with $\mathbf{S}_i$ being spin-$1/2$ operators. The spins are arranged on a square lattice with nearest-neighbor Heisenberg coupling $J'$. Pairs of spins form an effective square lattice of dimers (see Fig.~\ref{fig:SS}), with the Heisenberg coupling $J$ between spins within each dimer. The magnitude of the magnetic field is controlled by $h$. Below, we fix the units setting $J=1$ (as well as $k_B = \hbar = 1$).

At zero magnetic field, $h=0$, and for a small ratio of the couplings $J'/J$, the SSM has a dimer ground state, formed as a product of singlets~\cite{Shastry81}. For large $J'/J$, the ground state exhibits antiferromagnetic long-range order. In between, a plaquette phase is found~\cite{Koga00,Takushima01,Chung01,Laeuchli02} for $0.675(2) < J'/J < 0.765(15)$~\cite{Corboz13_shastry}.

At nonzero magnetic field, the model exhibits a series of magnetization plateaus. In this paper, we fix the ratio $J'/J=0.63$, which was estimated for SrCu$_2$(BO$_3$)$_2$ from fits to the magnetization curve at high fields, see Ref.~\onlinecite{matsuda13}. We simulate the model in the middle of the $m=1/2$ magnetization plateau for $h/J = 1.85$~\cite{matsuda13}. The ground state corresponding to this plateau breaks the translational symmetry and exhibits a checkerboard order of magnetized dimers.  The corresponding order parameter $o$ is  defined as
\be
o = \frac{1}{N} \left(\sum_{i \in A} \langle S^z_{i} \rangle - \sum_{i\in B} \langle S^z_{i} \rangle \right),
\ee
where the labels $A$ and $B$ distinguish dimers (and the spins forming them) belonging to two different checkerboard sublattices, see Fig.~\ref{fig:SS}. $N$ is the number of all dimers in the lattice.   The order parameter and the dimensionality of the model imply that a finite temperature second-order phase transition should belong to the universality class of the classical 2D Ising model.

\section{iPEPS algorithm}
\label{sec:iPEPS}

We use a FU scheme to simulate imaginary time evolution of a thermal state's purification that is represented as an iPEPS~\cite{CzarnikDziarmagaCorboz,CzarnikKH}. To simulate the SSM we map each dimer to a single iPEPS site with a local Hilbert space of dimension $d=4$, see Fig.~\ref{fig:SS} (as was performed in some of the previous studies~\cite{Corboz13_shastry,matsuda13,corboz14_shastry,boos19,wietek19}). We employ $U(1)$ symmetric iPEPS to speed up simulations. The expectation values of observables follow from iPEPS contraction via the corner transfer-matrix renormalization-group (CTMRG) method~\cite{Nishino_CTMRG_96, Orus_CTM_09, Corboz_CTM_14} where the accuracy of the contraction is controlled by an environmental bond dimension $\chi$.

For the largest simulated iPEPS bond dimension $D=9$, we find that it is beneficial to perform the first few steps of the evolution with the SU scheme and then continue with the FU scheme.
The imaginary time evolution begins at infinite temperature for which we conveniently choose the purification to be a product state over all the sites of the effective lattice of dimers,
\be 
\prod_k \left( \sum_{j_k=1}^d \left|j_k,j_k\right\rangle \right).
\ee
Here, the first (second) index in $\left|j_k,j_k\right\rangle$ refers to the physical (ancilla) state at the $k$-th lattice site. When represented by iPEPS the product state has a trivial bond dimension, $D=1$. 
In the early stages of the evolution, for small inverse temperatures $\beta$, correlations remain short-range and are effectively limited to the nearest-neighbor sites. In that case, it is not necessary to use the FU scheme that requires expensive evaluation of the infinite tensor environment (necessary to represent long-range correlations accurately) and whose stability may become problematic. Indeed, for weak correlations when $D$ is too large, the norm matrix (that has to be repeatedly pseudoinverted) is abundant in zero modes. 
We observe that the FU results for large $D$ depend strongly on the pseudoinverse cutoff chosen during the first steps of the evolution, consistent with this scenario.  Therefore, in the early stages of the evolution (in the high-temperature regime), the SU scheme offers more efficiency and stability than the FU, without compromising the accuracy of long-range correlations that are still absent. However, in the case of critical states, it was shown that the SU scheme might converge too slowly with the iPEPS bond dimension to provide accurate results~\cite{CzarnikDziarmagaCorboz}. That is why, as the temperature decreases towards $T_c$, the optimal strategy seems to be to switch from SU to FU at some inverse-temperature $\beta_{SU}$ chosen to maximize the accuracy (see Appendix~\ref{app:SU+FU} for systematic comparisons).  

We add a small symmetry-breaking term
\be
H_{\textrm{bias}} =  h_s \left( \sum_{i \in A} S_i^z - \sum_{i \in B} S_i^z \right),
\label{Hsim}
\ee
to make the simulations of the evolution across the critical point with a finite bond dimension feasible.
The bias $h_s$ turns the phase transition into a smooth crossover. Therefore, we perform simulations using the Hamiltonian,
\be
H = H_0  + H_{\textrm{bias}},
\ee
and recover the results for $H_0$ by extrapolating to $h_s=0$. To that end, we require a scaling theory in the critical regime. We begin with a simplifying assumption that the results are converged in the bond dimension for each value of the bias.

\section{Simple scaling theory}
\label{sec:simple}

Assuming a continuous phase transition---rather than a weakly first-order one---and convergence in $D$, the simple (standard) scaling theory predicts the behavior of the order parameter $o(t,h_s)$, its temperature derivative $o'(t,h_s) = \partial o(t, h_s)/\partial t$, specific-heat $C_V(t,h_s)$, and correlation length $\xi(t,h_s)$ in the vicinity of the critical temperature $T_c$ of the second-order phase transition as
\ba
o(t,h_s)  &=&  h_s^{1/\delta} f(t h_s^{-1/\tilde\beta\delta}),
\label{mscal} \\
o'(t,h_s) &=&  h_s^{(\tilde\beta-1)/\tilde\beta\delta} f'(t h_s^{-1/\tilde\beta\delta}),
\label{mderivscal} \\
C_V(t,h_s) &=&  h_s^{-\alpha/\tilde\beta\delta} g(t h_s^{-1/\tilde\beta\delta}),
\label{CVscal} \\
\xi(t,h_s) &=&  h_s^{-\nu/\tilde\beta\delta} l(t h_s^{-1/\tilde\beta\delta}). 
\label{xiscal} 
\ea
Here, $t = (T-T_c)/T_c$ is a dimensionless distance from the critical point, $\nu$, $\tilde\beta$, $\delta$, and $\alpha$ are the critical exponents, and $f$, $g$, and $l$ are nonuniversal functions with $f'(x)= df/dx$~\cite{CzarnikDziarmagaCorboz}. Note that we use $\tilde\beta$ instead of a conventional notation to avoid confusion with the inverse temperature $\beta = 1/T$. At the critical temperature (at $t=0$) the correlation length scales with $h_s$ as 
\be 
\xi(t=0,h_s) \defeq \xi_h \propto h_s^{-\nu/\tilde\beta\delta}.
\label{xih}
\ee 

\begin{figure}[t!]\centering
\includegraphics[width=0.98\columnwidth,clip=true]{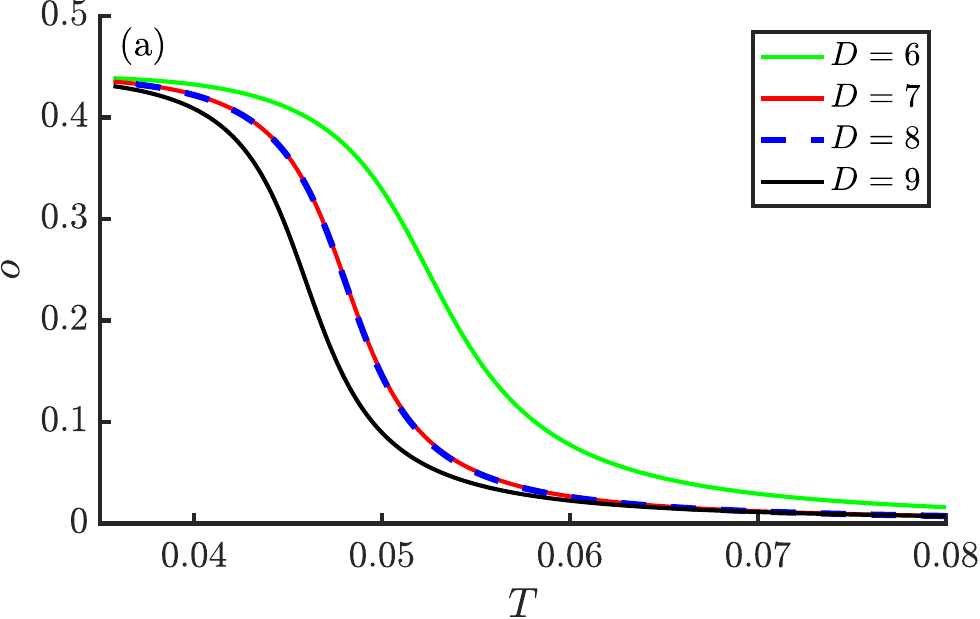}\vspace{0.2cm}
\includegraphics[width=0.98\columnwidth,clip=true]{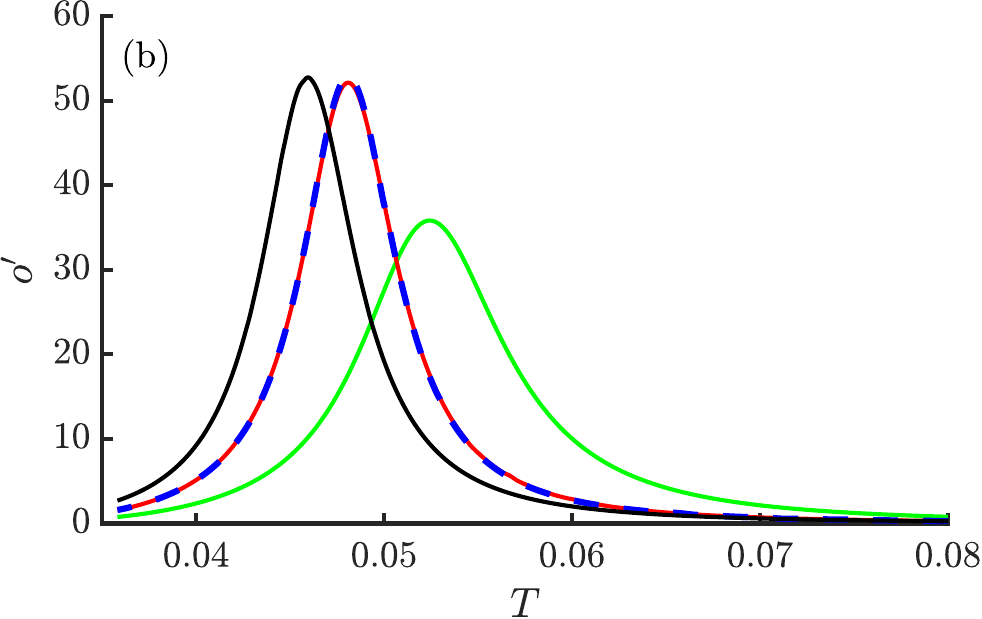}\vspace{0.2cm}
\includegraphics[width=1\columnwidth,clip=true]{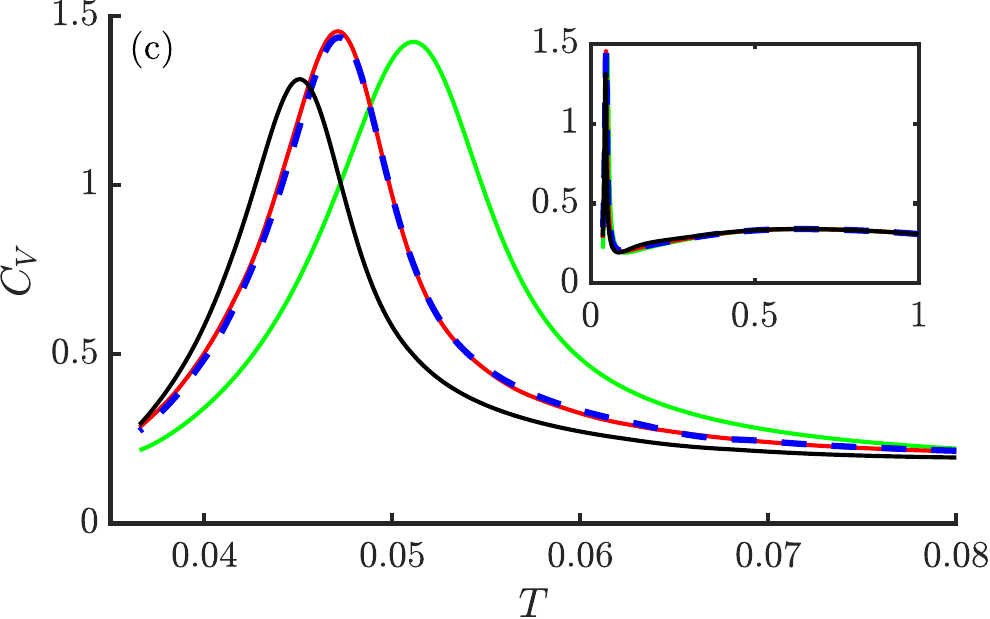}
\caption{
Temperature dependence of (a) the order parameter $o(T)$, (b) its temperature derivative $o'(T)$, and (c) the specific heat $C_V(T)$. The results are for the smallest simulated symmetry-breaking bias $h_s/h \approx 2.4 \times 10^{-4}$, and the parameters $J'/J = 0.63$ and $h/J=1.85$ (fixing the units with $J=\hbar=k_B=1$). Different curves correspond to the iPEPS bond dimension $D=6-9$, where the exact results should be recovered for $D\to\infty$.
Both $o'(T)$ and $C_V(T)$ display sharp peaks at temperatures where the order parameter suddenly rises, indicating the vicinity of the critical point. The inset displays a wider range of temperatures.
}
\label{fig:ord}
\end{figure}

\begin{figure}[t!] \centering
\includegraphics[width=1\columnwidth,clip=true]{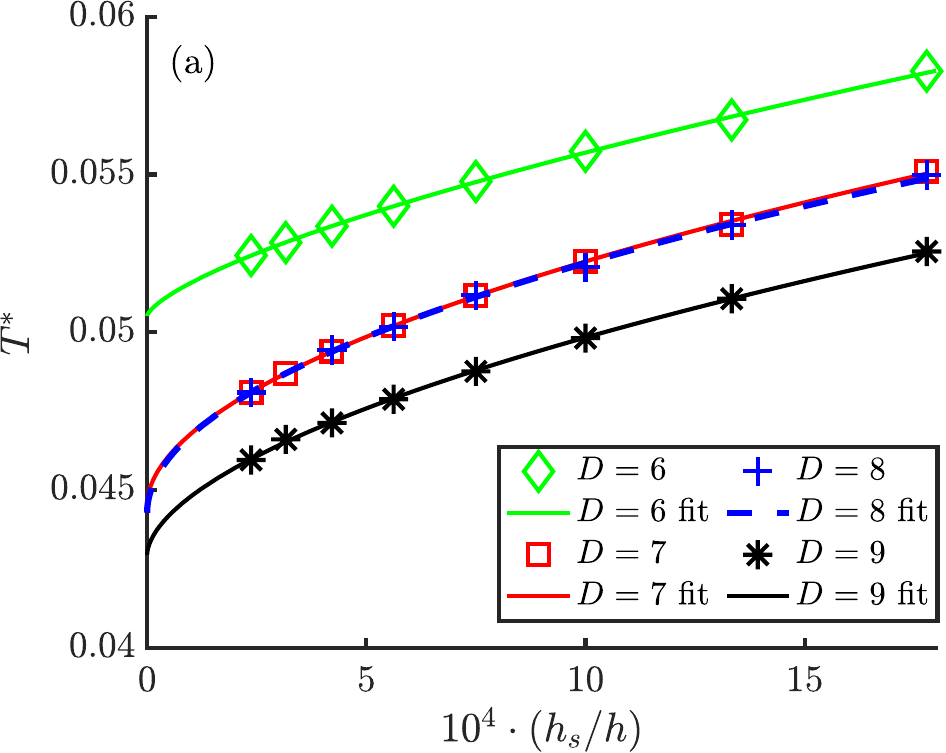} \vspace{0.2cm}
\includegraphics[width=1\columnwidth,clip=true]{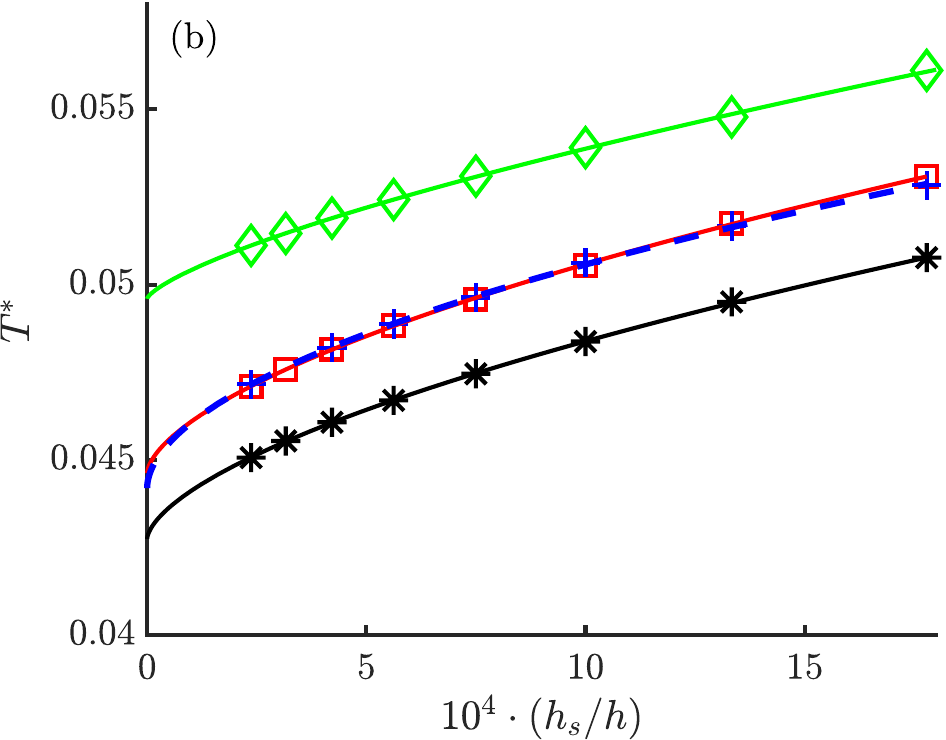}
\caption{Extrapolation of the critical temperature based on a simple scaling theory.
In panel (a), we plot the position $T^{*}(h_s)$ of the maximum of $o'(T, h_s)$, see Fig.~\ref{fig:ord}, for different values of the bias and $D=6-9$. Each curve is fitted with the scaling ansatz in Eq.~\eqref{Tast}. Similarly, in  (b), we focus on the maxima of the specific-heat $T^{*}_{C_V}(h_s)$ and fit it with the scaling ansatz in Eq.~\eqref{TastCV}. The obtained values of the critical temperature  $T_c$ and critical exponent $1/\tilde\beta\delta$ are collected in~Table~\ref{tab:Tc}.
}
\label{fig:Tast}
\end{figure}

For fixed $h_s$, both $o'(t,h_s)$ and $C_V(t,h_s)$ have a peak in the vicinity of $T_c$,  at $T^*(h_s)$ and $T^*_{C_V}(h_s)$, respectively. The scaling theory implies that
\ba  
T^*(h_s) &=& T_c + a h_s^{1/\tilde\beta\delta},
\label{Tast} \\
T_{C_V}^*(h_s) &=& T_c + b h_s^{1/\tilde\beta\delta},
\label{TastCV}
\ea
where $a$ and $b$ are nonuniversal constants. Below, we use Eqs.~\eqref{Tast} and \eqref{TastCV} to determine $T_c$ and verify the universality class numerically. Additionally, to provide further verification of the universality class, we investigate the behavior of the correlation length $\xi$ and order-parameter derivative $o'$ at $t^*(h_s)= [T^*(h_s)-T_c]/T_c$, i.e., at the temperature where $o'$ has a maximum. The expected scalings read
\ba
\xi^{*}(h_s) &\defeq& \xi(t^*(h_s),h_s) \propto  h_s^{-\nu/\tilde\beta\delta}, 
\label{xipeak} \\
o'^{*}(h_s) &\defeq& o'(t^*(h_s),h_s)  \propto  h_s^{(\tilde\beta-1)/\tilde\beta\delta}. 
\label{derivpeak}
\ea

\section{Numerical results}
\label{sec:results}

We perform the imaginary time evolution with a bias in the range $10^{-29/8} \approx 2.4\times10^{-4}  \le h_s/h \le 10^{-22/8} \approx 1.8\times 10^{-3}$ (with $h_s/h = 10^{-j/8}$ for integer $j$). Figure~\ref{fig:ord} shows the order parameter and the specific heat for the weakest bias, $h_s/h=10^{-29/8}$, where criticality is the most apparent. For $D=6-9$, we observe the order-parameter symmetry breaking and the peak of specific heat $C_V$ at $T\approx0.05$. Although it is clear that the results are not fully converged in the bond dimension, in this section we will continue with the simple scaling theory. A more refined analysis will follow in Sec.~\ref{sec:general}.

As we can see in Fig.~\ref{fig:Tast}, in the considered range of biases, $T^{*}(h_s)$ and $T^{*}_{C_V}(h_s)$ can be fitted accurately by scaling formulas in Eqs.~\eqref{Tast} and \eqref{TastCV}, respectively. From the fits, we infer the estimates of $T_c$ and $1/\tilde \beta \delta$ that we gather in Table~\ref{tab:Tc}. For the largest $D=7-9$, the estimates of $1/\tilde\beta\delta$ that we obtain from the fits of $T^*(h_s)$ agree with the classical 2D Ising universality class within their error bars. {\it A priori}, we expect that the estimates obtained from the fits of $T^{*}_{C_V}(h_s)$ may be less accurate as the height of the $C_V$ peak diverges only logarithmically in the 2D Ising universality class. That makes the scaling in Eq.~\eqref{TastCV} more susceptible to non-universal corrections. Nevertheless, for $D=7-9$, we still obtain $1/\tilde{\beta}\delta$ which differs from the Ising universality class by at most $15\%$~only.

\begin{table}[t]
\begin{tabular}{|c|c|c|c|l|l|}
\hline
{\rm Method} & $D$ & $T_c$  & $1/\tilde{\beta}\delta$ \\ 
\hline
$T^*(h_s)$ & $6$  & $0.0505(3)$  & $0.69(5)$  \\
$T^*(h_s)$  & $7$  & $0.0445(5)$  & $0.54(4)$  \\
$T^*(h_s)$  & $8$  & $0.0443(9)$  & $0.51(7)$   \\
$T^*(h_s)$ & $9$  & $0.0429(4)$  & $0.57(4)$   \\
$T^*_{C_V}(h_s)$ & $6$  & $0.0496(2)$  & $0.72(4)$  \\
$T^*_{C_V}(h_s)$ & $7$  & $0.0446(2)$  & $0.61(2)$  \\
$T^*_{C_V}(h_s)$ &$8$  & $0.0442(3)$  & $0.53(3)$   \\
$T^*_{C_V}(h_s)$ & $9$  & $0.0428(1)$  & $0.61(1)$ \\
{\rm 2D Ising}  &  & & $8/15 \approx  0.53$  \\
\hline
\end{tabular}
\caption{The critical temperature $T_c$ and the critical exponent $1/\tilde\beta\delta$ obtained from the best fits of $T^{*}(h_s)$ and $T^{*}_{C_V}(h_s)$ shown in Fig.~\ref{fig:Tast} for different values of the iPEPS bond dimension $D$. The error bars correspond to $68\%$ confidence intervals. The fitted values of $1/\tilde{\beta}\delta$ can be compared with the one in the universality class of the 2D classical Ising model. \hspace{-30pt}
}
\label{tab:Tc}
\end{table}

\begin{figure}[h]\centering
\includegraphics[width=0.96\columnwidth,clip=true]{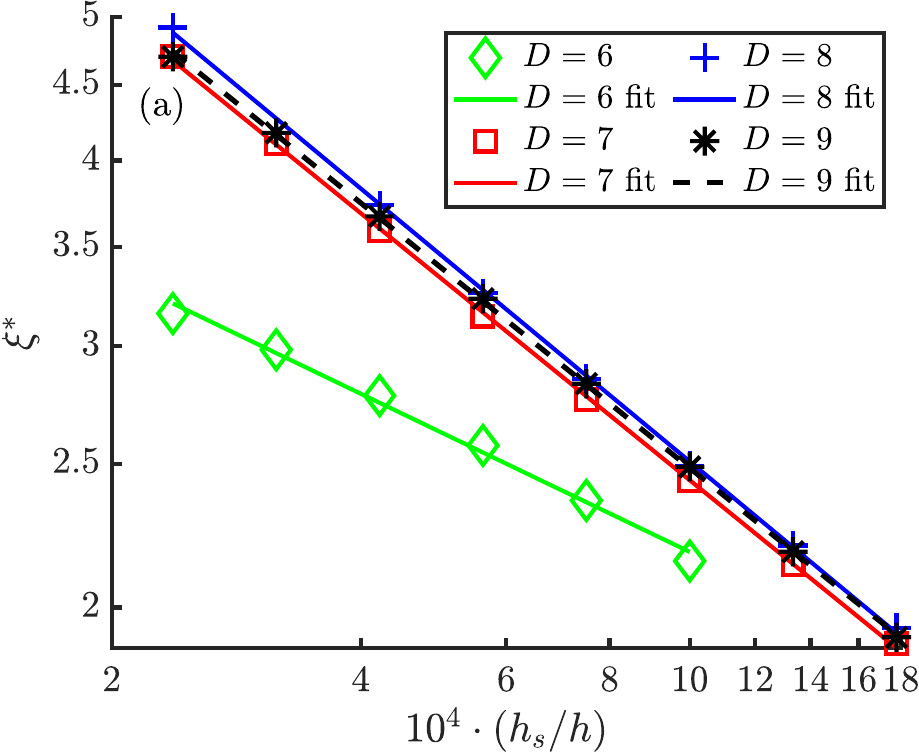}\\ \vspace{0.2cm}
\includegraphics[width=0.95\columnwidth,clip=true]{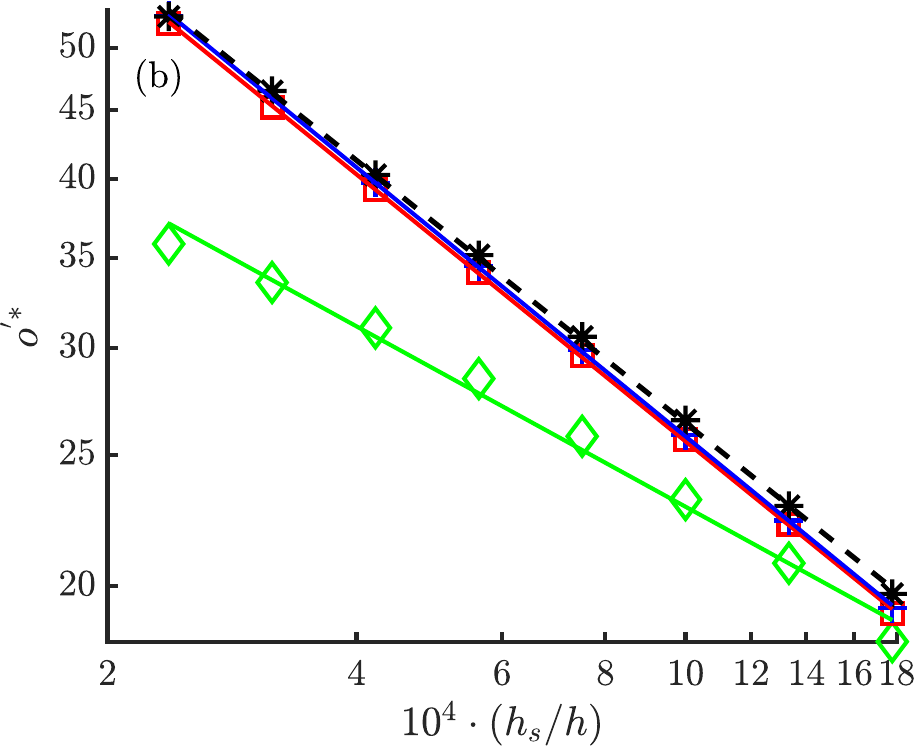}\\
\caption{Scaling of the correlation length and the magnitude of the order-parameter derivative at its peak. In (a), we show a log-log plot of $\xi^*(h_s)$ fitted by the scaling ansatz in Eq.~\eqref{xipeak}, whereas in (b), we focus on $o'^*(h_s)$ and the corresponding scaling ansatz in Eq.~\eqref{derivpeak}. The fitted critical exponents are gathered in Table~\ref{tab:loglogxi}. Here $\xi^*$ is measured in units of lattice spacing of the effective dimer lattice, see Fig.~\ref{fig:SS}.
}
\label{fig:loglogxi}
\end{figure}

We proceed with analyzing the magnitudes of the correlation length and the order-parameter derivative. In Fig.~\ref{fig:loglogxi} we plot $o'^{*}(h_s)$ and $\xi^*(h_s)$ for $D=7-9$ together with their best fits according to the scaling formulas \eqref{derivpeak} and \eqref{xipeak}, respectively. Estimates of $(1-\tilde \beta)/\tilde \beta \delta$ and $\nu/\tilde\beta\delta$ obtained in this way are collected in Table~\ref{tab:loglogxi}. They differ from the Ising 2D values by at most $7\%$ and $15\%$, respectively.
Here, the correlation length was obtained by extrapolating the spectrum of the CTMRG transfer matrix  to the limit of infinite environmental bond dimension, $\chi$~\cite{Rams_xiD_18}. We obtain
\be 
1.9 < \xi^{*}(h_s) < 4.9
\ee
measured in the units of the dimer lattice spacing (for the units of spacing in the original spin-$1/2$ lattice, multiply by $\sqrt{2}$). The values compare favorably to the ones accessible by other state-of-the-art methods, such as DMRG on a thin cylinder.

\begin{table}[h!]
\begin{tabular}{|c|c|c|c|l|l|}
\hline
 $D$    & $\nu/\tilde\beta\delta$  &  $(1-\tilde\beta)/\tilde\beta\delta$\\ 
\hline
$6$   & $0.27(2)$  & $0.33(2)$ \\
 $7$   & $0.452(3)$ & $0.496(2)$  \\
$8$   & $0.462(5)$ & $0.500(3)$   \\
$9$  & $0.449(3)$ &  $0.488(4)$   \\
2D Ising & $8/15 \approx  0.53$ & $7/15 \approx  0.467$ \\
\hline
\end{tabular}
\caption{ The values of the critical exponents $\nu/\tilde\beta\delta$ and $(1-\tilde\beta)/\tilde\beta\delta$ obtained from the best fits of the correlation length and the peak amplitude of the order-parameter derivative in Fig.~\ref{fig:loglogxi} for different iPEPS bond dimensions $D$. The reference values are for the classical 2D Ising universality class.}
\label{tab:loglogxi}
\end{table}

\section{General scaling theory}
\label{sec:general}

Although our numerical simulation could not be fully converged in $D$ in the vicinity of the transition, the simple scaling theory---which assumes such a convergence---provides remarkably self-consistent results. Those are in good agreement with the expected universality class of the 2D Ising model. This suggests that we are in the limit where the corrections to the simple scaling theory are small. To capture them correctly, we try below a more general scaling theory. There, in addition to the correlation length $\xi_h$ in \eqref{xih} (that diverges in the limit of zero bias), we have another correlation length $\xi_D$ that diverges for infinite bond dimension. Here, we define $\xi_D$ as the correlation length obtained for a given $D$ at the critical point (at $t=0$) and without bias $h_s=0$. In the absence of the bias, the relevance of $\xi_D$ was well established both at zero~\cite{Corboz_FCLS_18,Rader_FCLS_18} and at finite temperatures~\cite{czarnik19b}. 

When both $\xi_h$ and $\xi_D$ are finite, Eq.~\eqref{mderivscal} generalizes to 
\be
o'(t,h_s, D) = h_s^{(\tilde{\beta}-1)/\tilde{\beta}\delta} F(t h_s^{-1/\tilde{\beta}\delta}, \xi_h/\xi_D),
\label{mderivscalunif}
\ee
where $F$ is a non-universal function. For $\xi_h/\xi_D\to0$, it becomes equal to $f'(t h_s^{-1/\tilde{\beta}\delta})$ in Eq.~ \eqref{mderivscal}. From Eq.~\eqref{mderivscalunif} we obtain 
\be
T^*(h_s, D) = T_c + a(x) h_s^{1/\tilde \beta \delta}, \quad x = \xi_h/\xi_D,
\label{Tastscalunif}
\ee
where $a(x)$ is a non-universal function. We are interested in the regime of $x \ll 1$, where the influence of the bias $h_s$ on the correlation length is stronger than the one coming from $D$ (in other words, results for each finite bias are almost converged in $D$). In this regime we postulate an asymptotic form
\begin{equation}
a(x) = a + bx^c
\label{a}
\end{equation}
where $a$, $b$, and $c$ are non-universal constants. The combination of Eqs.~\eqref{Tastscalunif} and \eqref{a} yields 
\begin{equation}
T^*(h_s,\xi_D) = T_c + a h_s^{1/\tilde \beta \delta} + \frac{b}{\xi_D^c} h_s^{(1-c\nu)/\tilde\beta  \delta}.
\label{Tastscalunif2}
\end{equation}
Here $T_c$, $a$, $b$, and $c$ are the fitting parameters that do not depend on the bond dimension, and $\xi_D$ is $D$ dependent. Note that it is not possible to fit the actual values of $\xi_D$ but only the ratios $\xi_{D+1}/\xi_D$. 

\begin{figure}[t!]
\includegraphics[width=1\columnwidth,clip=true]{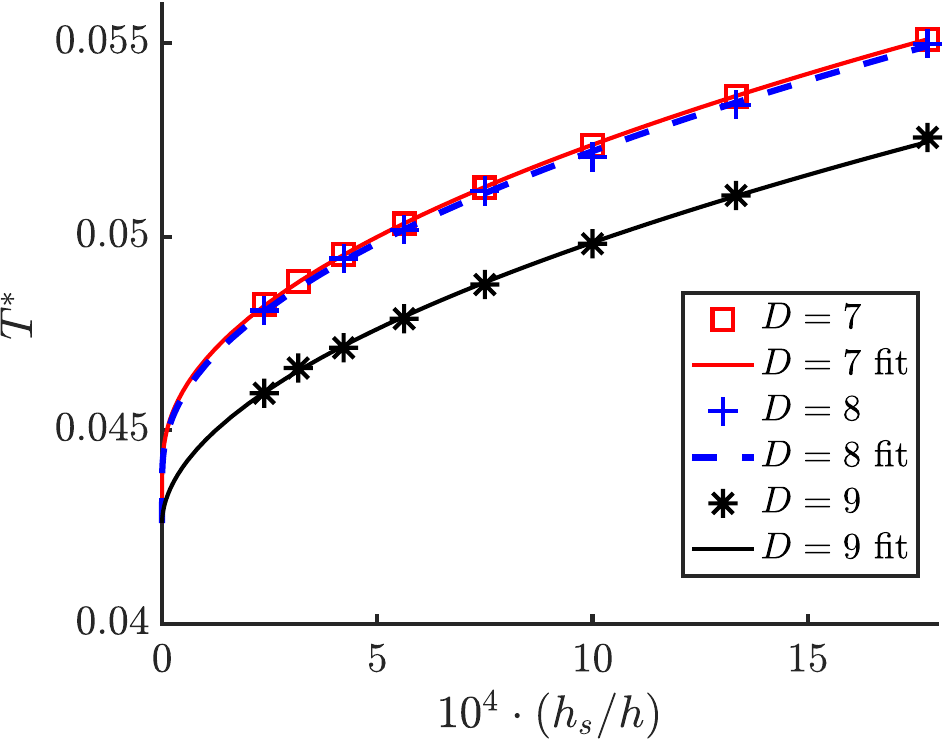}\\
\caption{The scaling ansatz combining both the effect of a small bias $h_s$ and the finite iPEPS bond dimension $D$. All three curves $T^{*}(h_s, \xi_D)$, indicating the position of the maximum of the order parameter derivative for $D=7,8,9$, are fitted by the single scaling ansatz in Eq.~\eqref{Tastscalunif2}. The fit yields $T_c=0.043(2)$, $c=0.83(4)$, $\xi_{D=8}/\xi_{D=7}=1.09(7)$, $\xi_{D=9}/\xi_{D=7}=25^{+\infty}_{-23.6}$. The error bars correspond to $68\%$ confidence intervals.  
}
\label{fig:Tast_univ}
\end{figure}

Assuming the 2D Ising universality class, the best fit to all the data for $D=7,8,9$ is shown in Fig.~\ref{fig:Tast_univ}. It yields the critical temperature
\be 
T_c=0.043(2),
\label{Tc}
\ee 
with the error bar corresponding to a $68\%$ confidence interval. We note that the upper error bar of $\xi_{D=9}/\xi_{D=7}$ extends to infinity.
This is a consequence of $(1-c\nu)/\tilde\beta\delta=0.09(2)$ being close to zero which makes the last term in \eqref{Tastscalunif2} rather insensitive to $h_s$. For $1-c\nu=0$ it would be impossible to fit $T_c$. This explains its relatively large error bar, despite the high quality of the fit in Fig.~\ref{fig:Tast_univ}. 
Still, the relative error on $T_c$ is only $5\%$, which is a considerably high accuracy given the fact that the error includes both finite bias and finite bond dimension effects in a systematic way.

\section{Summary}
\label{sec:summary}
In this paper, we have systematically studied the finite temperature phase transition into the $m=1/2$ plateau phase in the Shastry-Sutherland model using iPEPS. 
At the technical level there are two main developments in this paper: the SU + FU hybrid algorithm and the generalized scaling theory. The former combines the advantages of the two most popular time-evolution schemes. At a small $\beta$, it takes advantage of the stability and efficiency of the SU for states with short-range correlations. At larger $\beta$'s it switches over to the FU scheme to take into account long-range correlations and make the most efficient use of the limited bond dimension. The latter development becomes a necessary tool when even the FU algorithm cannot provide full convergence in the vicinity of the phase transition for available bond dimensions. The generalized scaling theory includes two relevant finite length scales: One due to the finite symmetry-breaking bias and the other due to the finite bond dimension. Its expressive power bridges the gap towards results that are fully converged in both bond dimension and extrapolated to zero bias.  

Using the simple scaling theory, which assumes results are converged in $D$, the estimates of $1/\tilde\beta\delta$ obtained with the scaling ansatz~\eqref{Tast} are in agreement with the 2D classical Ising model universality class for sufficiently large $D\ge7$. Our result for the critical temperature $T_c=0.043(2)$, obtained from the generalized scaling theory, is compatible with the observation of a stable $m=1/2$ plateau in experiments at a temperature around $2.1$~K~\cite{matsuda13}. Taking the estimate of $J=84$~K from fits to  specific heat data~\cite{wietek19}, we obtain a critical temperature $T_c=3.5(2)$~K, which is well above the temperature used in the experiment. It would be interesting to study the stability of the $m=1/2$ plateau as a function of temperature in future experiments.

\acknowledgements
We acknowledge funding by Narodowe Centrum Nauki (NCN), Poland under Project No. 2019/35/B/ST3/01028 (J.D. and M.M.R.), the Laboratory Directed Research and Development (LDRD) program of Los Alamos National Laboratory (LANL) under Project No. 20190659PRD4 (P.C.). This project has received funding from
the European Research Council (ERC) under the European Union’s Horizon 2020 Research and Innovation
Programme (Grant Agreement No. 677061). This work is part of the D-ITP consortium, a program of the Netherlands Organization for Scientific Research (NWO) that is funded by the Dutch Ministry of Education, Culture and Science (OCW).

\appendix

\section{The simple update to the full update (SU + FU) and the FU approaches---technical details}
\label{app:SU+FU}

\begin{figure}[t!] \centering
\includegraphics[width=0.95\columnwidth,clip=true]{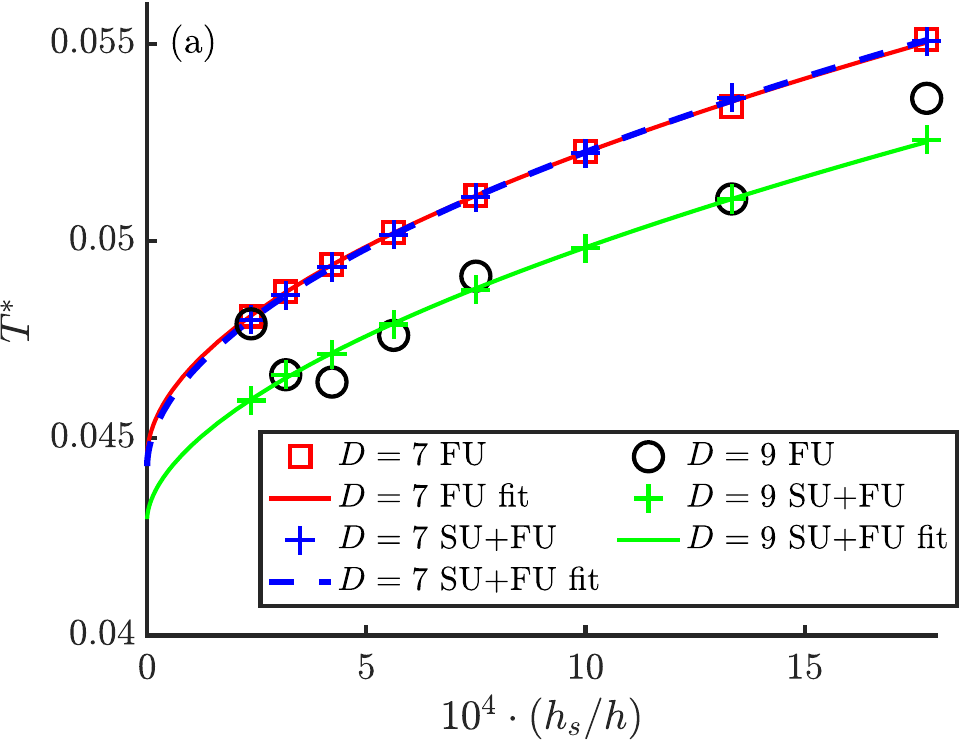}\\ \vspace{0.2cm}
\includegraphics[width=0.95\columnwidth,clip=true]{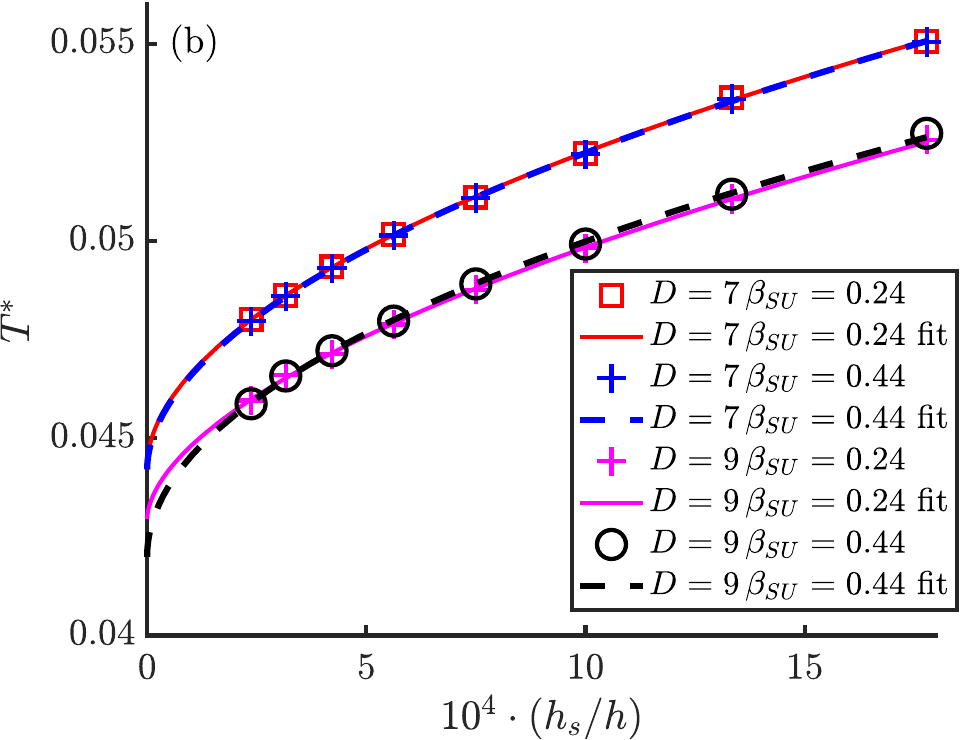}\\
\caption{A comparison of FU and FU+SU approaches. In (a), we show $T^{*}(h_s/h)$ where the SU + FU results were obtained with $\beta_{SU}=0.24$. In (b) we show the comparison of SU + FU results obtained with $\beta_{SU}=0.24$ and $\beta_{SU}=0.44$. 
We observe that for $D=7$ both FU and SU + FU give similar results, but for $D=9$ SU + FU gives more regular results. For both $D=7$ and $9$, $\beta_{SU}=0.24$ and $0.44$ give similar results. The fitted $T_c$ and $1/\tilde\beta\delta$ can be found in Table~\ref{tab:TcSUFU}. }
\label{fig:TastSUFU}
\end{figure}

Within the FU approach, for the largest $D=9$, we found that the results depend on simulation parameters such as the magnitude of the pseudoinverse cutoff. The same is true for charges and dimensions of the U(1) symmetry sectors of the iPEPS tensors (which are selected dynamically during the simulation). Furthermore, for $D=9$, we observed different iPEPS tensors' decomposition into symmetry sectors for different small $h_s$, even when using the same FU pseudoinverse cutoff. A change in charge distribution can lead to a qualitative change in the data, leading to an irregular behavior that cannot be explained by a simple scaling theory. Indeed, we observe that for $D=9$,  $T^*(h_s)$ obtained with FU is not monotonic, see Fig.~\ref{fig:TastSUFU}(a). The same figure shows that performing the first few steps of the evolution with the SU solves the problem. Namely, the iPEPS tensors' charge distributions become the same for all small values of $h_s$'s, and $T^*(h_s)$ becomes monotonic in $h_s$. 

To obtain a stable charge sector decomposition, for $D=9$, it is enough to perform the SU evolution until $\beta_{SU}=0.24$. $D=9$ results shown in the main text have been obtained with this $\beta_{SU}$.  We find that in the case of smaller $D=7-8$, performing the first steps with the SU is not necessary. In the case of $D=7$ and $h_s$ considered in the main text, iPEPS tensors have the same decomposition. For $D=8$ we have found just one $h_s$ with a different decomposition than the rest, which we have excluded from the analysis. 

\begin{table}[h!]
\begin{tabular}{|c|c|c|c|l|l|}
\hline
{\rm Method} & $D$ & $T_c$  & $1/\tilde\beta\delta$ \\ 
\hline
FU & $7$  & $0.0445(5)$  & $0.54(4)$  \\
SU + FU $\beta_{SU}=0.24$& $7$  & $0.0443(3)$  & $0.53(2)$   \\
SU + FU $\beta_{SU}=0.44$& $7$  & $0.0442(4)$  & $0.53(2)$   \\
SU + FU $\beta_{SU}=0.24$& $9$  & $0.0429(4)$  & $0.57(4)$   \\
SU + FU $\beta_{SU}=0.44$& $9$  & $0.0420(6)$  & $0.50(4)$   \\
2D Ising &  & & $8/15 \approx  0.53$  \\
\hline
\end{tabular}
\caption{Comparison of critical temperature $T_c$ and exponent  $1/\tilde\beta\delta$ obtained from the best fits in Fig.~\ref{fig:TastSUFU}. }
\label{tab:TcSUFU}
\end{table}

Nevertheless, for a benchmark purpose and cross-check, we simulate $D=7$ with the SU + FU method using $\beta_{SU}=0.24$ and $0.44$. We find that the obtained results are very similar to the ones coming from the FU method, see Fig.~\ref{fig:TastSUFU}(b) and Table~\ref{tab:TcSUFU}. To get a better insight into the stability of the results as a function of $\beta_{SU}$, we also simulate $D=9$ with $\beta_{SU}=0.44$. With this choice, we obtain iPEPS tensors with a different charge distribution than with $\beta_{SU}=0.24$, but the estimates of $T_c$ and $1/\tilde\beta\delta$ are still within their respective error bars for both $\beta_{SU}$ values, see Table~\ref{tab:TcSUFU}. 

In view of the increased stability of the FU+SU scheme, one may wonder if it would be beneficial to switch to the cheaper SU scheme completely. To give insight into this question, in the next appendix, we compare the SU + FU results with the pure SU approach.

The computational complexity of our full update evolution  implementation  is determined by sizes and numbers of  $U(1)$-symmetric sectors, see Appendix~\ref{app:sectors} for their definitions. Therefore, scaling of  the complexity with increasing $D$ can not be determined {\it a priori}. For the largest simulated $D=9$ simulations took about 2 weeks to complete using a  machine with 4 Intel Xeon E7-8890v4 24-core processors. 

\begin{figure}[t!] \centering
\includegraphics[width=0.95\columnwidth,clip=true]{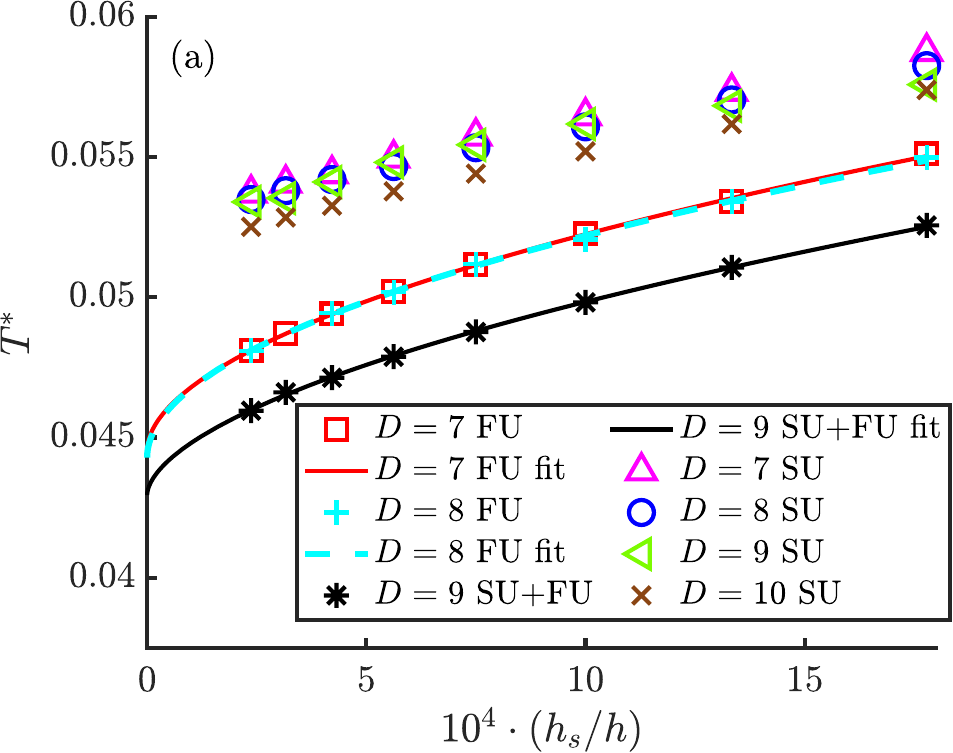}\\ \vspace{0.2cm}
\includegraphics[width=0.95\columnwidth,clip=true]{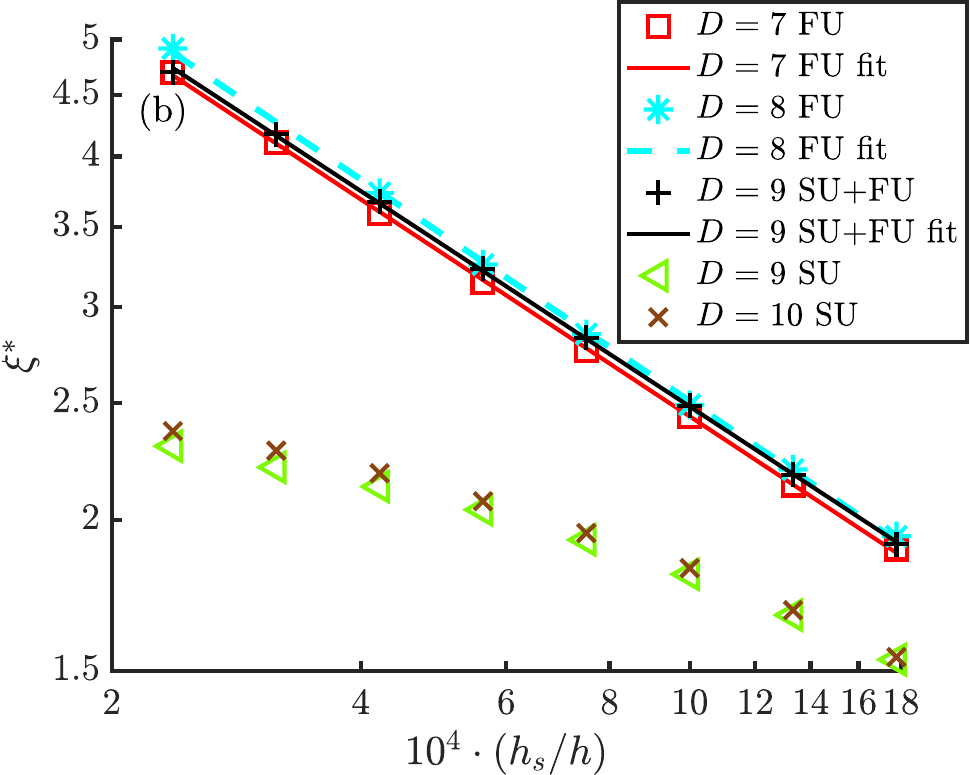}
\caption{ 
Comparison of the SU approach with the FU and FU+SU results. In (a)  we show $T^*(h_s/h)$, where the SU results have been obtained with $D=7-10$. They differ significantly from the FU and SU + FU results, changing slowly with increasing $D$. In (b) we show a log-log plot of the correlation length $\xi^*(h_s/h)$.  The SU results for $D=9-10$ are clearly inconsistent with the expected power-law behavior of $\xi^*(h_s/h)$.
}
\label{fig:SU}
\end{figure}

\section{Comparison with the simple update (SU) approach}
\label{app:SU}

Here, we compare the SU approach results for $D=7-10$ with the FU and SU + FU at $D=7-9$. We find that $T^*(h_s)$ obtained with SU are far removed from the FU results and changing slowly with increasing $D$, see Fig.~\ref{fig:SU}(a). In Fig.~\ref{fig:SU}(b), we show a log-log plot of the SU $D=9-10$ results for $\xi^*(h_s)$ together with their FU and SU + FU counterparts. They deviate significantly from the critical behavior. We also note that correlation lengths $\xi^*(h_s)$ obtained from the SU are much shorter than those obtained from the FU and SU + FU. We conclude that the convergence of the SU results in $D$ is much slower than in the case of the FU and SU + FU methods. Therefore,  we prefer to use the FU and SU + FU here. 

\section{Effects of finite CTMRG environmental bond dimension $\chi$ and Trotter step $d\beta$}
\label{app:chidbeta}

\begin{table}[b!]
\begin{tabular}{|c|c|c|c|l|l|}
\hline
$\chi$ & $d\beta$ & $T_c$  & $1/\tilde\beta\delta$ \\ 
\hline
$21$ & $0.04$  & $0.0445(5)$  & $0.54(4)$  \\
$21$ & $0.02$  & $0.04470(3)$  & $0.53(2)$  \\
$28$ & $0.04$  & $0.0444(4)$  & $0.53(3)$  \\
$35$ & $0.04$  & $0.0442(3)$  & $0.53(3)$  \\
\hline
\end{tabular}
\caption{Convergence of critical temperature $T_c$ and exponent $1/\tilde\beta\delta$ with Trotter step $d\beta$ and environmental CTMRG bond dimension $\chi$. The results are obtained from the best fits of $T^{*}(h_s)$ for the FU scheme with $D=7$.}
\label{tab:Tcchi}
\end{table}

\begin{table*}[t!]
\begin{tabular}{|c|c|c|c|c|c|}
\hline
$D$ & $(c_t,D_{c_t})$ & $(c_l,D_{c_l})$  & $(c_b,D_{c_b})$  & $(c_r, D_{c_r})$ & $(c_i,d_{c_i})$  \\ 
\hline
$6$ & $\{(-1,2),(0,2),(1,2)\}$  & $\{(-1,2),(0,3),(1,1)\}$  & $\{(-1,1),(0,3),(1,2)\}$ & $\{(-1,1),(0,3),(1,2)\}$ & $\{(0,1),(1,2),(2,1)\}$ \\
$7$ & $\{(-1,2),(0,3),(1,2)\}$  & $\{(-1,2),(0,3),(1,2)\}$  & $\{(-1,2),(0,3),(1,2)\}$ & $\{(-1,2),(0,3),(1,2)\}$ & $\{(0,1),(1,2),(2,1)\}$ \\
$8$ & $\{(-1,3),(0,3),(1,2)\}$  & $\{(-1,3),(0,3),(1,2)\}$  & $\{(-1,2),(0,3),(1,3)\}$ & $\{(-1,2),(0,3),(1,3)\}$ & $\{(0,1),(1,2),(2,1)\}$ \\
$9$ & $\{(-1,3),(0,4),(1,2)\}$  & $\{(-1,3),(0,4),(1,2)\}$  & $\{(-1,2),(0,4),(1,3)\}$ & $\{(-1,2),(0,4),(1,3)\}$ & $\{(0,1),(1,2),(2,1)\}$ \\
\hline
\end{tabular}
\caption{Charges of the $A^{A}$ sectors and dimensions of the sectors' indices for $D=6-9$ simulations from the main text and large $\beta>0.32$. We show the decomposition for the smallest $h_s=10^{-29/8}$.  Here we  denote bond dimensions of $A^{c_t,c_l,c_b,c_r,c_i,c_j}$ by $D_{c_t}, D_{c_l}, D_{c_b}, D_{c_r}$ and a dimension of its physical index by $d_{c_i}$.
Charges and dimensions for the ancillary index $j$ are the same as for the physical index $i$.
Note that an analogous decomposition of $A^B$ is determined by the $A^{A}$ decomposition.   }
\label{tab:sec}
\end{table*}

The CTMRG environmental bond dimension $\chi$  controls the accuracy of the iPEPS contraction. To perform the FU evolution and to obtain $o$ , $o'$ and $C_V$ shown in the main text, we use $\chi = 3D-4D$.  Another parameter that determines the accuracy of the simulation is the Trotter step $d\beta$. In the main text, we use a second-order Trotter decomposition with the step $d\beta=0.04$. Here we compare the $T_c$ and $1/\tilde\beta$ estimates obtained from $T^*(h_s)$ following the FU evolution with $D=7, d\beta=0.04$ and $\chi=21,28,35$. Furthermore, we compare the results obtained with $d\beta=0.02, 0.04$ and $D=7,\chi=21$. We find that the results are very similar, see Table~\ref{tab:Tcchi}, which suggests that the chosen values of $d\beta$ and $\chi$ are good enough to provide accurate estimates. To obtain the correlation length $\xi$, we perform the extrapolation following the approach of Ref.~\cite{Rams_xiD_18} using the data from the range of environmental bond dimensions of $\chi =D^2-3D^2$.

\section {A PEPS tensors' decomposition to $U(1)$-symmetric sectors}
\label{app:sectors}

The Hamiltonian  conserves $S^z_{tot}$. Therefore, the thermal density matrix $\rho$ is $U(1)$ invariant. To create a $U(1)$-invariant tensor network representation of $\rho$, we choose the following representation of the symmetry group
\begin{equation}
U\rho U^{\dag} = \rho, \quad U = \otimes U^{(d)}, \quad U^{(d)}=e^{-i(S^z_1+S^z_2+1)},
\end{equation}
where the tensor product is taken over the dimers and $S^z_1$, $S^z_2$ are acting at the first and the second spins of a dimer, respectively. $\rho$ is built from a pair of PEPS tensors $A^A$ and $A^B$ corresponding to dimer sub-lattices $A$ and $B$. To simplify notation, we omit the sub-lattice index below whenever possible.   The  PEPS tensors $A_{tlbrij}$ have six indices, where $t$, $l$, $b$, and $r$ are virtual indices, $i$ is a physical index corresponding to a dimer and $j$ is an ancilla index. We choose the tensors to be $U(1)$ invariant,
\begin{equation}
\sum_{t'l'b'r'i'j'}U_{tt'}^{(t)} U^{(l)}_{ll'} U^{(b)\dag}_{bb'} U^{(r)\dag}_{rr'}  U^{(d)}_{ii'}   U_{jj'}^{(d)\dag} A_{t'l'b'r'i'j'} = A_{tlbrij}.
 \label{Asymm}
\end{equation}
Here we introduce  group representations $U^{(l)}$, $U^{(b)}$, $U^{(r)}$, and $U^{(t)}$ acting at virtual indices  as proposed in Refs.~\cite{singh2010-1,bauer2011}. 

 For such a choice,  $A$  can be decomposed into symmetric sectors~\cite{singh2010-1} indexed by integer charges $c_t$, $c_l$, $c_b$, $c_r$, $c_i$, and $c_j$,
 \begin{equation}
 A = \oplus_{ c_t,c_l,c_b,c_r,c_i,c_j}
 A^{c_t,c_l,c_b,c_r,c_i,c_j}, 
 \label{dec}
\end{equation}
with a constraint  
 \begin{equation}
  c_t+c_l-c_b-c_r+c_i-c_j=0.
\label{constraint}  
 \end{equation}
In Table~\ref{tab:sec}, we list the charges and dimensions of the sectors for $D=6-9$ simulations.
We observe that for our $D=6-9$ simulations described in the main text charges present in  the $A$ decomposition (\ref{dec}) remain  unchanged for $\beta > \beta_{SU}$ (or $\beta>0$ in the case of full update simulations). 
All combinations of the charges allowed by (\ref{constraint}) are present  for $\beta > 0.04$.
Furthermore,  dimensions of the charge sectors' indices remain unchanged  for $\beta>0.32$.

%


\end{document}